\documentclass[preprint,aps,showpacs,showkeys,dvips]{revtex4}

\usepackage{graphicx}
\usepackage{dcolumn}
\usepackage{bm}
\usepackage{subfigure}

\begin{document}

\title{Anomalous Single Production of the Fourth Generation Quarks at the LHC}

\author{R. \c{C}ift\c{c}i}
\email{rena.ciftci@cern.ch}
\affiliation{Dept. of Eng. of Physics, Faculty of Eng., Ankara University, 06100 Tandogan,
Ankara, Turkey}

\date{\today}

\begin{abstract}
Possible anomalous single productions of the fourth standard model generation up and down type quarks at CERN Large Hadron Collider are studied. Namely, $p p \rightarrow u_4 (d_4) X$ with subsequent  $u_4\rightarrow bW^{+}$ process followed by leptonic decay of $W$ boson and $d_4\rightarrow b\gamma$ (and its h.c.) decay channel are considered. Signatures of these processes  and corresponding standard model backgrounds are discussed in detail. Discovery limits for quark mass and achievable values of anomalous coupling strength are determined.
\end{abstract}
\maketitle
\section{Introduction}	

The number of fundamental fermion generations is not predicted by the Standard Model (SM). This number is restricted from below with LEP I data on invisible decays of Z boson as $n_g\geq3$ \cite{Yao06}. On the other hand, the asymptotic freedom of QCD enforces the number of generations to be less than 9. The recent precision electroweak data are equally consistent with existence of three or four SM generations \cite{He01,Novikov02,Kribs07}.

The flavor democracy is a natural hypothesis in the framework of SM as well as a number of models dealing with new physics (see review \cite{Sultansoy07} and references therein). The flavor democracy hypothesis in SM predicts the existence of the fourth SM generation \cite{Fritzsch87,Datta93,Celikel95}. Tevatron experiments give $250$ GeV of lower bound on the mass of the fourth generation quarks \cite{CDF07}. However it does not fix the masses and mixings of the new fermions. Another prediction of the flavor democracy is the masses of the fourth generation fermions to be nearly degenerate and lie between 300 and 700 GeV, whereas, the masses of known fermions belonging to lighter three generations appear due to small deviations of the democracy \cite{Datta94,Atag96,Ciftci05}. The last value is close to the upper limit on heavy quark masses, which follows from partial-wave unitary at high energies \cite{Chanowitz79}. The quark masses and Cabibbo Kobayashi Maskawa (CKM) matrix for certain parametrization of mass matrix are given in Refs. \cite{Datta94,Atag96}. Ref. \cite{Ciftci05} gives both masses and CKM matrix (Maki Nakagawa Sakata (MNS) matrix for leptons) for both quarks and leptons for another parametrization.

\begin{figure}
\subfigure[]{\includegraphics{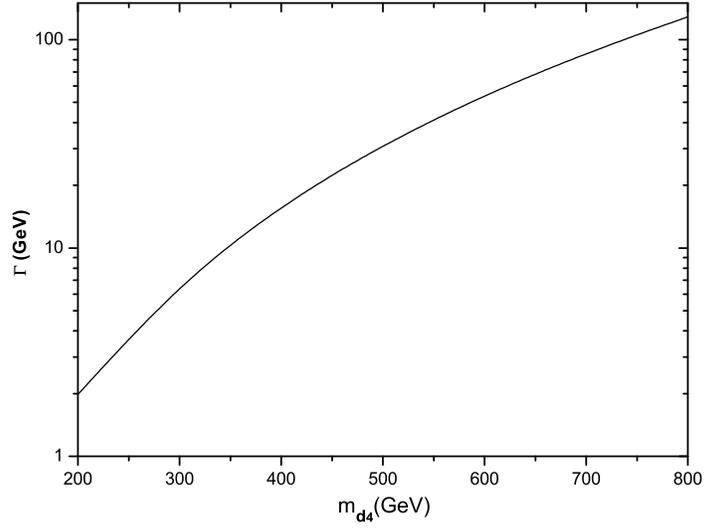}}
\subfigure[]{\includegraphics{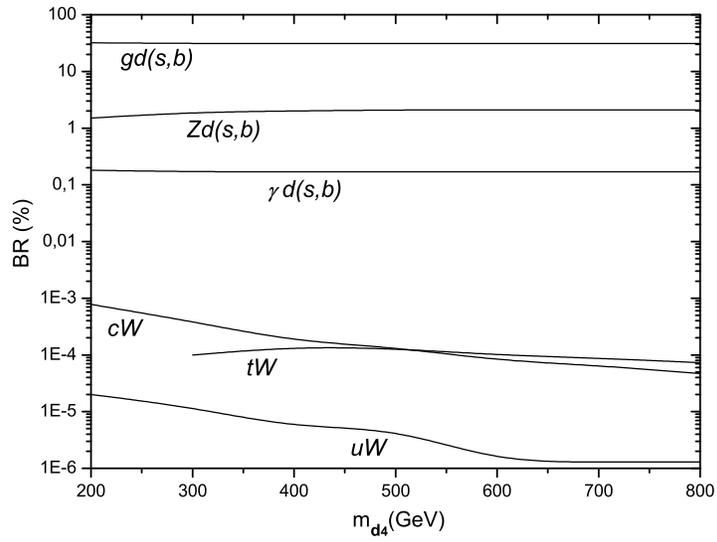}}
\caption{(a) The total decay width of the fourth generation down type quark and (b) the branching ratios (\%) depending on the mass of the quark. \protect\label{fig1}}
\end{figure}

A new era in High Energy Physics will be opened with the construction of TeV scale colliders \cite{Sultansoy98}. Obviously, TeV energy colliders are needed for discovery of the fourth SM generation fermions. The first of the such colliders is the Large Hadron Collider (LHC) with 14 TeV center of mass energy. The fourth generation quarks is predicted to be copiously produced in pairs at the LHC \cite{ATLAS99,Arik99}. Recently, this process is proposed as the best scenario (after Higgs) for discovery at the LHC \cite{Holdom06,Holdom07a,Holdom07b}. Also, a serious contributions of anomalous interactions for the production of the fourth generation fermions can be expected. Such anomalous interactions seems to be quite natural due to large masses of the fourth generation fermions (see argumentation for anomalous interaction for t quark presented in ref. \cite{Fritzsch99}). These anomalous interactions could provide also single production of the fourth generation fermions. The anomalous single production of the fourth generation fermions is considered in a number of papers \cite{Arik03a,Arik03b,Ciftci07a,Ciftci07b}. Recently, anomalous production of the fourth generation charged lepton and neutrino at future ep colliders is considered in \cite{Ciftci08a} and \cite{Ciftci08b}, respectively. This paper is devoted to study the anomalous single production of the fourth generation up and down type quarks at the LHC. 

In section II the Lagrangian for SM and the anomalous interactions of the fourth generation quarks is presented; the decay width and branching ratios of the fourth generation quarks are evaluated. Productions of the fourth generation quarks at the LHC are studied in section III: $pp\rightarrow d_4X\rightarrow b\gamma X$ (and its h.c.) and $pp\rightarrow u_4X\rightarrow b W^{+}X \rightarrow b\nu \ell^{+} X$ processes as well as their SM backgrounds are considered. The statistical significance of the signal and achievable values of anomalous coupling strength are given with concluding remarks in Section IV.

\section{Anomalous Interactions of the Fourth Generation Quarks}

The effective Lagrangian for the anomalous interactions of $u_4$ and $d_4$ quarks can be rewritten from \cite{Rizzo97,Arik03b} with minor modifications as:

\begin{equation}
{\cal L}=\left(\frac{\kappa^{q_{i}}_{\gamma}}{\Lambda}\right) e_{q}g_{e}\bar{q}_{4}\sigma_{\mu\nu}q_{i} F^{\mu\nu}+\left(\frac{\kappa^{q_{i}}_{Z}}{2\Lambda}\right) g_{Z}\bar{q}_{4}\sigma_{\mu\nu}q_{i} Z^{\mu\nu}+\left(\frac{\kappa^{q_{i}}_{g}}{\Lambda}\right) g_{s}\bar{q}_{4}\sigma_{\mu\nu}T^{a}q_{i} G^{\mu\nu}_{a}+h.c. \hspace{2mm}  ,  \hspace{3mm} 
\end{equation}
where $\textit{i}=1,2,3$ denotes the generation index. $\kappa^{q_{i}}_{\gamma}$, $\kappa^{q_{i}}_{Z}$ and $\kappa^{q_{i}}_{g}$  are the anomalous couplings for the electromagnetic, weak (neutral current) and strong interactions, respectively (in numerical calculations, $\kappa^{q_{i}}_{\gamma}= \kappa^{q_{i}}_{Z} = \kappa^{q_{i}}_{g}$ is assumed). $\Lambda$ is the cutoff scale for the new physics and $e_q$ is the quark charge. $g_e$, $g_Z$ and $g_s$ are the electroweak and strong coupling constants. In the above equation, $\sigma_{\mu\nu} = i(\gamma_{\mu}\gamma_{\nu}-\gamma_{\nu}\gamma_{\mu})/2$. $F^{\mu\nu}$, $Z^{\mu\nu}$ and $G^{\mu\nu}_{a}$ are the field strength tensors of the photon, Z boson and gluons, respectively. $T_a$ is the Gell-Mann matrices.

\begin{figure}
\subfigure[]{\includegraphics{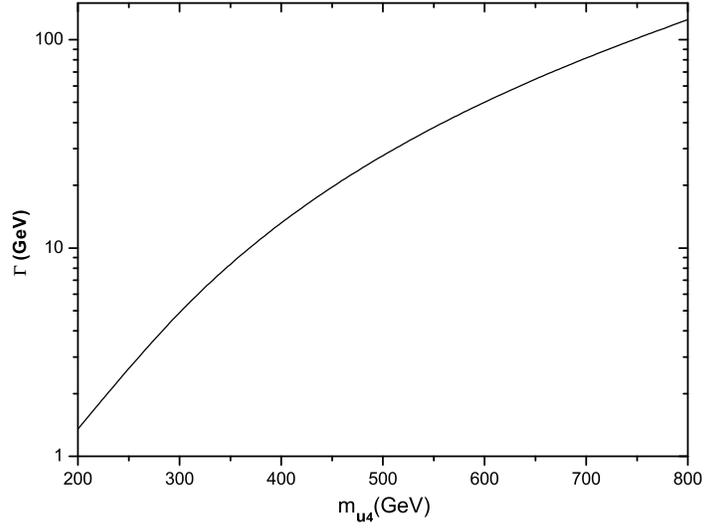}}
\subfigure[]{\includegraphics{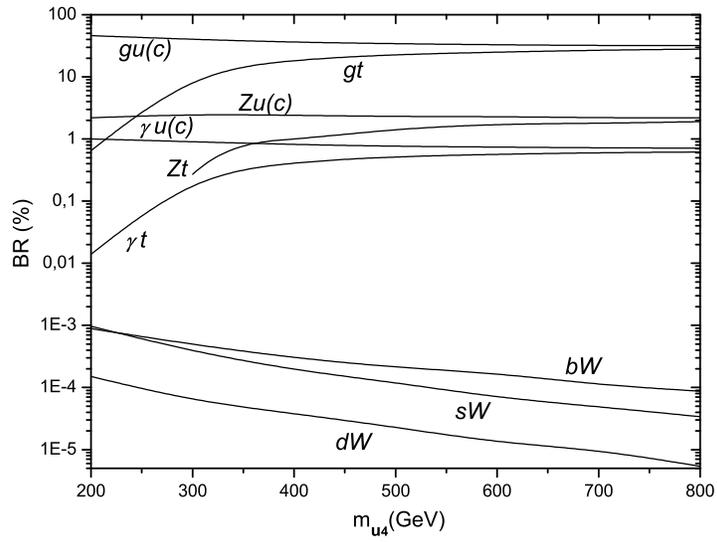}}
\caption{(a) The total decay width of the fourth generation up type quark and (b) the branching ratios (\%) depending on the mass of the quark. \protect\label{fig2}}
\end{figure}

\begin{figure}
\includegraphics{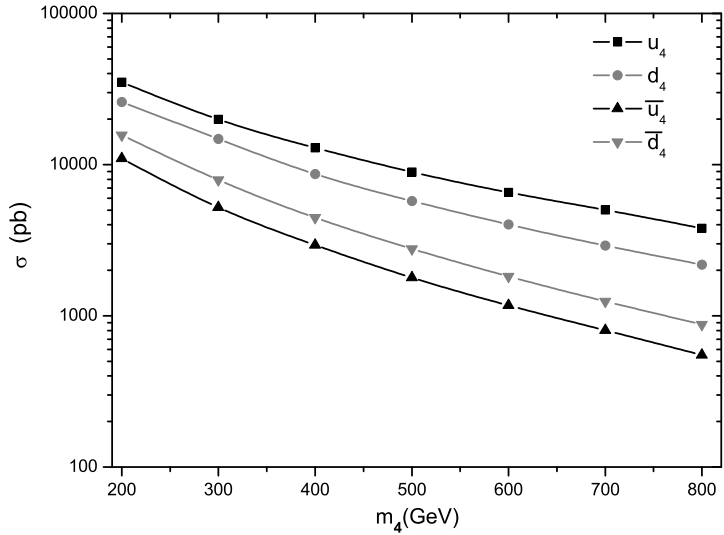}
\caption{The total production cross-sections of the $p p \rightarrow u_4 (\bar{u_4}) X$ and $p p \rightarrow d_4 (\bar{d_4}) X$ processes at the LHC. \protect\label{fig3}}
\end{figure}

\begin{figure}
\subfigure[]{\includegraphics{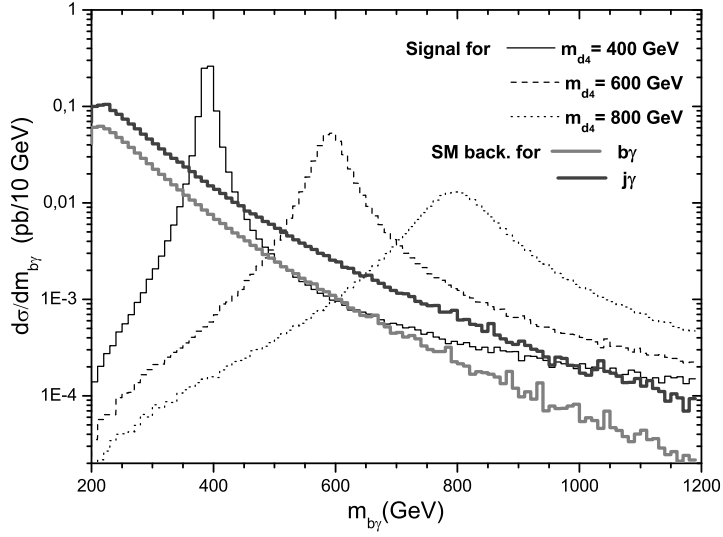}}
\subfigure[]{\includegraphics{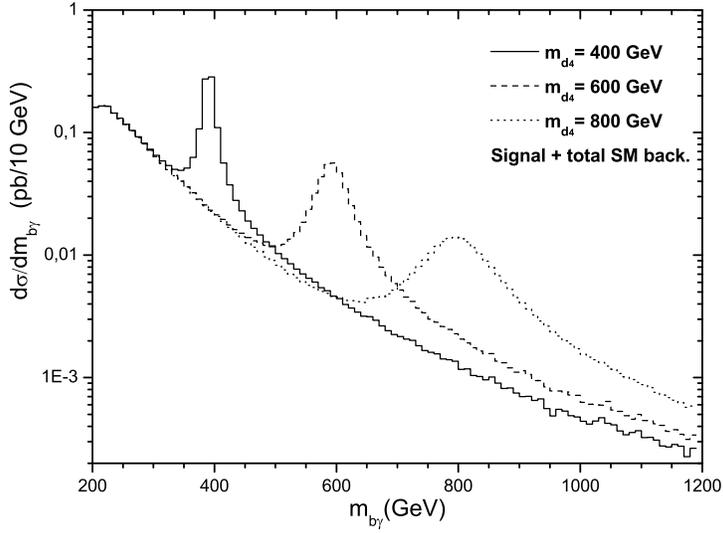}}
\caption{Invariant mass distributions of the fourth generation down type quark signal for three selected mass values and corresponding SM backgrounds at the LHC (with cut Selection3 given in Table I): (a) separated and (b) summed.\protect\label{fig4}}
\end{figure}

\begin{table}
\caption{Cut selection criteria and corresponding signal and SM background cross sections for $pp\rightarrow d_4X\rightarrow b\gamma X$ process (and its h.c.) (at $(\kappa/\Lambda)$ = 1TeV$^{-1}$)}
\begin{ruledtabular}
\begin{tabular}{cccc}
 &\multicolumn{3}{c} {Signal cross sections (pb)} \\
\cline{2-4}
 & \textbf{Selection1} & \textbf{Selection2} & \textbf{Selection3}  \\ 
 & $P^{\gamma}_{T}\:>\:20$ GeV & $P^{\gamma}_{T}\:>\:20$ GeV, $P^{j}_{T}\:>\:100$ GeV,  & Selection2 + \\
$m_{d_4}$(GeV)&  & $\left|\eta_{j,\gamma}\right|\:<\:2.5$, $\Delta R_{j,\gamma}\:>\:0.4$ & b-tag + mistag \\
\hline
300 & 38.54 & 22.89 & 12.82\\
400 & 22.26 & 16.12 & 9.03\\
500 & 14.49 & 11.36 & 6.36\\
600 & 9.93 & 8.19 & 4.58\\
700 & 7.08 & 6.03 & 3.38\\
800 & 5.19 & 4.52 & 2.53\\
\hline
 Process &\multicolumn{3}{c} {SM background cross sections (pb)}  \\ 
\hline
$pp\rightarrow b\gamma X+h.c.$ & $5.60\times 10^3$  & $1.25\times 10^1$  & 7.01\\
$pp\rightarrow c\gamma X+h.c.$ & $3.42\times 10^4$ & $7.96\times 10^1$ &  7.96\\
$pp\rightarrow j_{light}\gamma X$ & $1.58\times 10^5$ & $2.83\times 10^2$ & 2.83 \\
Total & $1.98\times 10^5$ & $3.75\times 10^2$ & 17.80 \\
\end{tabular}
\end{ruledtabular}
\end{table}
 
Obviously new interactions will lead to additional decay channels of the fourth generation quarks. In order to compute decay widths, I have implemented the new interaction vertices into the CompHEP \cite{Boos04}. While calculating the SM decay contributions, CKM mixings given in Ref. \cite{Ciftci05} are used. However, CKM Matrices are not given excluding the fourth generation quark mass values of 400 GeV and 800 GeV in Ref. \cite{Ciftci05}. Therefore, CKM matrix elements for other mass values are obtained by using the same parametrization to be consistent on the rest of calculations. SM decay widths are proportional to ${\left| V_{q_4 q^{'}}\right|}^{2}$. Depending on relative magnitudes of $(\kappa/\Lambda)$ and $\left| V_{q_4 q^{'}}\right|$, SM or anomalous decays will dominate. The total decay width $\Gamma$ of the fourth generation quarks and the relative branching ratios are presented in Fig. 1 and Fig. 2 for $(\kappa/\Lambda)$ = 1 TeV$^{-1}$. Concerning the anomalous coupling strength, the value $(\kappa/\Lambda)$ = 1 TeV$^{-1}$ is rather conservative (the mass scale is order of electroweak scale). Consequently, I have used this value at rest of my calculations.

\section{Anomalous Single Production of the Fourth Generation Quarks at LHC}

I have calculated the anomalous single production cross sections of the fourth SM generation quarks at the LHC using CompHEP with CTEQ6L1 parton distribution functions \cite{Pumplin02}. Contributions to single production of $u_4$ come from $gu$ and $gc$ processes. Similar contributions for $pp\rightarrow d_4 X$ come from processes of $gd$, $gs$ and $gb$. The results are shown in Fig. 3 for $(\kappa/\Lambda)$ = 1 TeV$^{-1}$.

Various signatures for anomalous interactions of the fourth generation quarks might be considered. One can group these signatures as
\begin{itemize}
	\item Anomalous production followed by anomalous decay such as $pp\rightarrow d_4 X\rightarrow g b X$, $pp\rightarrow d_4 X\rightarrow ZbX$, $pp\rightarrow d_4 X \rightarrow \gamma bX$, $pp\rightarrow u_4 X\rightarrow g tX$ and $pp\rightarrow u_4 X\rightarrow \gamma tX$
	\item Anomalous production followed by SM decay such as $pp\rightarrow u_4 X\rightarrow b W^+ X$ and $pp\rightarrow d_4 X \rightarrow t W^- X$.
\end{itemize}
In this study, one process is chosen from each group of signatures. Namely, $pp\rightarrow d_4 X \rightarrow \gamma bX$ from "anomalous production-anomalous decay" group and $pp\rightarrow u_4 X\rightarrow b W^+ X$ from "anomalous production-SM decay" group are chosen to investigate in details because of  their relative simplicity. While the cross section of the first group is proportional with $(\kappa/\Lambda)^4$, the cross section of the second group is proportional with  $(\kappa/\Lambda)^2$ and $\left| V_{u_4 b}\right|^2$. Therefore, values of $(\kappa/\Lambda)$ and $\left| V_{u_4 b}\right|$ determine which group of signature is important for the observatibility.

\begin{figure}
\subfigure[]{\includegraphics{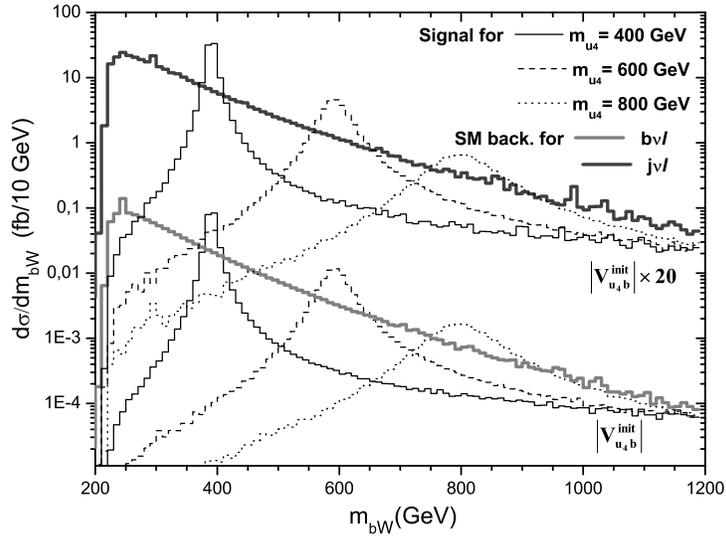}}
\subfigure[]{\includegraphics{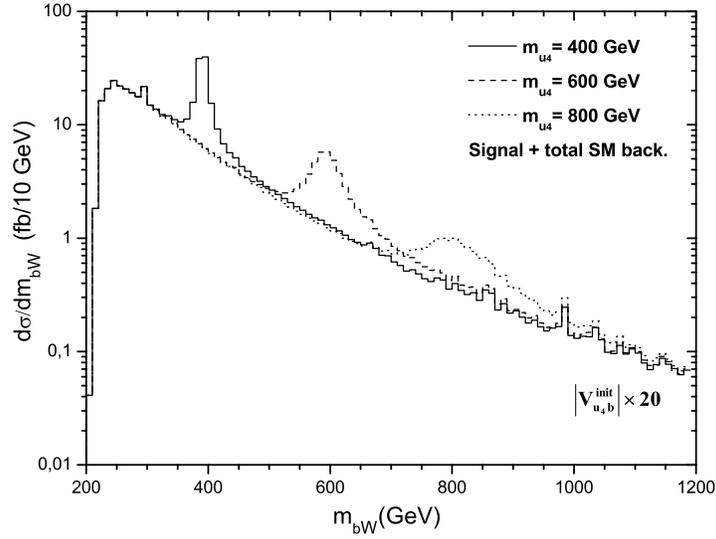}}
\caption{Invariant mass distributions of the fourth generation up type quark signal for some mass and $\left|V_{{u_4}b}\right|$ values and corresponding SM backgrounds at the LHC (with cut Selection1 given in Table II and $b$-tag + mistag ): (a) separated and (b) summed.\protect\label{fig5}}
\end{figure}

First, $p p \rightarrow d_4 X\rightarrow b\gamma X$ process (and its h.c.) is considered as signature of anomalous interactions of the fourth generation down type quark. The SM background for this process is potentially much larger than the signal. However, some kinematic cuts have been applied in order to extract the signal and to suppress the SM background. The following selection of cuts is chosen: $P_T\:>\:20$ GeV for photon, $P_T\:>\:100$ GeV for all jets; $\left|\eta_{j,\gamma}\right|\:<\:2.5$, where $\eta$ denotes pseudorapidity; a minimum separation of $\Delta R\:=\:\left[\left(\Delta\phi\right)^{2}+\left(\Delta\eta\right)^{2}\right]^{1/2}\:>\:0.4$ ($\phi$ is the azimuthal angle) between the photon and jets. In addition, I assume that $b$-quark jet is tagged with efficiency of $56\%$. Moreover, $1\%$ of light jets (for $u$, $d$, $s$, $\bar{u}$, $\bar{d}$, $\bar{s}$ and $g$) and $10\%$ of $c$-quark jets are mistagged as b-jet. The calculated signal and SM background cross sections before and after all these cuts, are given in Table I. Figure 4 shows the invariant mass distribution for the signal and the SM background for sample values of the fourth generation quark masses of $400$, $600$ and $800$ GeV.

\begin{table}
\caption{Cut selection criteria for $pp\rightarrow u_4X\rightarrow b\nu \ell^{+} X$ process}
\begin{ruledtabular}
\begin{tabular}{lll}
\textbf{Selection1}&\textbf{Selection2}&\textbf{Selection3}\\
\hline
$P^{\ell}_{T}\:>\:10GeV$               &Selection1+                                                  &Selection2+   \\
$P^{j}_{T}\:>\:100GeV$                 &$\left|M_{j\nu \ell}-M_{u_4}\right|\:<\:2\Gamma_{Tot}        $&b-tag+  \\
$\left|\eta_{j,\ell}\right|\:<\:2.5$   &                                                             &mistag \\
$ \Delta R_{j,\ell}\:>\:0.4$           &                                                             &       \\
\end{tabular}
\end{ruledtabular}
\end{table}

\begin{table}
\caption{Signal and SM background cross sections for $pp\rightarrow u_4X\rightarrow b\nu \ell^{+} X$ process with cut selections given in Table II (at $(\kappa/\Lambda)$ = 1TeV$^{-1}$)}
\begin{ruledtabular}
\begin{tabular}{lclcccccc}
\multicolumn{2}{l|} {Selection}& &\multicolumn{6}{c} {Signal cross sections (pb)}  \\
\cline{4-9}
\multicolumn{2}{l|} {Criterion} & $\left| V_{u_4 b}\right|$ & 300 GeV & 400 GeV& 500 GeV& 600 GeV & 700 GeV& 800 GeV  \\ 
\hline
\multicolumn{2}{l|} {\textbf{Selection1}} & $6 \cdot\left| V^{init}_{u_4 b}\right|$ &$1.36\cdot10^{-1}$& $8.71\cdot10^{-2}$ &$4.82\cdot10^{-2}$&$3.11\cdot10^{-2}$&$1.65\cdot10^{-2}$&$1.03\cdot10^{-2}$  \\ 
\multicolumn{2}{l|} {}              & $10\cdot\left| V^{init}_{u_4 b}\right|$ &$4.64\cdot10^{-1}$& $2.43\cdot10^{-1}$ &$1.34\cdot10^{-1}$&$8.69\cdot10^{-2}$&$4.04\cdot10^{-2}$&$2.86\cdot10^{-2}$  \\
\multicolumn{2}{l|} {}              & $20\cdot\left| V^{init}_{u_4 b}\right|$ &$1.86$& $9.71\cdot10^{-1}$ &$5.36\cdot10^{-1}$&$3.48\cdot10^{-1}$&$1.83\cdot10^{-1}$&$1.15\cdot10^{-1}$ \\
\multicolumn{2}{c|} {Process }  &                                       & \multicolumn{6}{c} {SM background cross sections(pb) }  \\
\hline
 \multicolumn{2}{l|} {$pp\rightarrow  b\nu \ell^{+} X$ }        & &\multicolumn{6}{c} {0.23 }  \\
 \multicolumn{2}{l|} {$pp\rightarrow  j_{light}\nu \ell^{+} X$ }& &\multicolumn{6}{c} {$1.31\cdot 10^{3}$ }  \\
 \multicolumn{2}{l|} {$pp\rightarrow  c\nu \ell^{+} X$ }        & &\multicolumn{6}{c} {$1.69\cdot 10^{2}$ }  \\
 \multicolumn{2}{l|} {Total }                               & &\multicolumn{6}{c} {$1.48\cdot 10^{3}$ }  \\
 \hline
\multicolumn{2}{c|} {}& &\multicolumn{6}{c} {Signal cross sections (pb) }  \\
\hline
 \multicolumn{2}{l|} {\textbf{Selection2}} & $6\cdot\left| V^{init}_{u_4 b}\right|$ &$1.11\cdot10^{-1}$& $7.25\cdot10^{-2}$ &$4.00\cdot10^{-2}$&$2.56\cdot10^{-2}$&$1.36\cdot10^{-2}$&$8.67\cdot10^{-3}$  \\ 
\multicolumn{2}{c|} {}               & $10\cdot\left| V^{init}_{u_4 b}\right|$ &$3.78\cdot10^{-1}$& $2.01\cdot10^{-1}$ &$1.11\cdot10^{-1}$&$7.15\cdot10^{-2}$&$3.79\cdot10^{-2}$&$2.41\cdot10^{-2}$  \\
\multicolumn{2}{c|} {}               & $20\cdot\left| V^{init}_{u_4 b}\right|$ &$1.51$& $8.07\cdot10^{-1}$ &$4.44\cdot10^{-1}$&$2.86\cdot10^{-1}$&$1.52\cdot10^{-1}$&$9.64\cdot10^{-2}$  \\
\multicolumn{2}{c|} {Process}               &                              & \multicolumn{6}{c} {SM background cross sections(pb) }  \\
\hline
 \multicolumn{2}{l|} {$pp\rightarrow  b\nu \ell^{+} X$ }        & &$1.11\cdot10^{-3}$& $1.17\cdot10^{-3}$ &$9.16\cdot10^{-4}$&$6.95\cdot10^{-4}$&$5.89\cdot10^{-4}$&$5.51\cdot10^{-4}$  \\
 \multicolumn{2}{l|} {$pp\rightarrow  j_{light}\nu \ell^{+} X$ }& &7.31& 9.18 &8.38&7.19&6.75&6.64   \\
 \multicolumn{2}{l|} {$pp\rightarrow  c\nu \ell^{+} X$ }        & &0.96& 1.04 &0.83&0.64&0.55&0.51   \\
 \multicolumn{2}{l|} {Total }                               & &8.27& 10.22 &9.22&7.83&7.30&7.15   \\
  \hline
\multicolumn{2}{c|} {}& &\multicolumn{6}{c} {Signal cross sections (pb) }  \\
\hline
 \multicolumn{2}{l|} {\textbf{Selection3}}& $6\cdot\left| V^{init}_{u_4 b}\right|$ &$6.21\cdot10^{-2}$& $4.06\cdot10^{-2}$ &$2.24\cdot10^{-2}$&$1.44\cdot10^{-2}$&$7.63\cdot10^{-3}$&$4.86\cdot10^{-3}$  \\ 
\multicolumn{2}{c|} {}              & $10\cdot\left| V^{init}_{u_4 b}\right|$ &$2.12\cdot10^{-1}$& $1.13\cdot10^{-1}$ &$6.22\cdot10^{-2}$&$4.00\cdot10^{-2}$&$2.12\cdot10^{-2}$&$1.35\cdot10^{-2}$  \\
\multicolumn{2}{c|} {}              & $20\cdot\left| V^{init}_{u_4 b}\right|$ &$8.47\cdot10^{-1}$& $4.52\cdot10^{-1}$ &$2.49\cdot10^{-1}$&$1.60\cdot10^{-1}$&$8.49\cdot10^{-2}$&$5.40\cdot10^{-2}$  \\
\multicolumn{2}{c|} {Process}              &                                         & \multicolumn{6}{c} {SM background cross sections(pb) }  \\
\hline
 \multicolumn{2}{l|} {$pp\rightarrow  b\nu \ell^{+} X$ }        & &$6.22\cdot10^{-4}$& $6.55\cdot10^{-4}$ &$5.13\cdot10^{-4}$&$3.89\cdot10^{-4}$&$3.30\cdot10^{-4}$&$3.10\cdot10^{-4}$  \\
 \multicolumn{2}{l|} {$pp\rightarrow  j_{light}\nu \ell^{+} X$ }& &$7.31\cdot10^{-2}$& $9.18\cdot10^{-2}$ &$8.38\cdot10^{-2}$&$7.19\cdot10^{-2}$&$6.75\cdot10^{-2}$&$6.64\cdot10^{-2}$ \\
 \multicolumn{2}{l|} {$pp\rightarrow  c\nu \ell^{+} X$ }        & &$9.62\cdot10^{-2}$& $1.04\cdot10^{-1}$ &$8.35\cdot10^{-2}$&$6.38\cdot10^{-2}$&$5.48\cdot10^{-2}$&$5.47\cdot10^{-2}$   \\
 \multicolumn{2}{l|} {Total }                               & &$1.70\cdot10^{-1}$& $1.97\cdot10^{-1}$ &$1.68\cdot10^{-1}$&$1.36\cdot10^{-1}$&$1.23\cdot10^{-1}$&$1.22\cdot10^{-1}$   \\
\end{tabular}
\end{ruledtabular}
\end{table}

\begin{table}
\caption{Statistical significances (SS) for anomalous interactions of the fourth generation up and down type quarks at the LHC with integrated luminosity of 100 fb$^{-1}$ }
\begin{ruledtabular}
\begin{tabular}{ccccc}
&$SS$ for $pp\rightarrow d_4X\rightarrow b\gamma X$& \multicolumn{3}{c} {$SS$ for $pp\rightarrow u_4X\rightarrow b\nu \ell^{+} X$}  \\
\cline{3-5}
$m_{4}$ (GeV) & &$6\cdot\left| V^{init}_{u_4 b}\right|$ &$10\cdot\left| V^{init}_{u_4 b}\right|$ & $20\cdot\left| V^{init}_{u_4 b}\right|$ \\ 
\hline
300\hphantom{00} & \hphantom{0}960 & \hphantom{0} 82.3  & \hphantom{0}230 & 920 \\
400\hphantom{00} & \hphantom{0}677 & \hphantom{0} 41.0  & \hphantom{0}114 & 456 \\
500\hphantom{00} & \hphantom{0}477 & \hphantom{0} 24.4  & \hphantom{0}68  & 272 \\
600\hphantom{00} & \hphantom{0}343 & \hphantom{0} 17.6  & \hphantom{0}49  & 196 \\
700\hphantom{00} & \hphantom{0}253 & \hphantom{0} 9.7   & \hphantom{0}27  & 108 \\
800\hphantom{00} & \hphantom{0}190 & \hphantom{0} 6.2   & \hphantom{0}17  & 69  \\
\end{tabular}
\end{ruledtabular}
\end{table}

Second I study $p p \rightarrow u_4 X\rightarrow bW^{+}X$ process at the LHC as signature of anomalous interactions of the fourth generation up type quark, followed by leptonic decay of $W^+$ boson ($\ell = e^+,\:\mu^+$). Computations show that the signal cross section for Selection3 given in Table II spans between 2.12 fb for $m_4=300$ GeV and 0.13 fb for $m_4=800$ GeV. Naturally, these values are too small with respect to the background as can be seen from Fig. 5. Main reason for the low values of the signal cross section is the initial mixing value between $u_4$ and $b$ given by parametrization used. In the parametrization, $\left| V^{init}_{u_4 b}\right|$ takes values from 0.0017 for $m_4=300$ GeV to 0.0008 for $m_4=800$ GeV. Without experimental data it is not possible to know the correct CKM matrix. Therefore, $\left|V_{{u_4}b}\right|$  might be bigger than $\left| V^{init}_{u_4 b}\right|$ of the parametrization. As earlier mentioned, SM decays are dependent on CKM matrix element. In this paper, the signal cross section values are computed with $\left| V_{u_4 b}\right|=\left| V^{init}_{u_4 b}\right|\times 6$, $\left| V^{init}_{u_4 b}\right|\times 10$ and $\left| V^{init}_{u_4 b}\right|\times 20$ for illustration and given in Table III for different cut selection criteria given in Table II.

The SM background for this process is studied in detail.  In order to extract the signal and to suppress the SM background, some kinematic cuts are tried. These cut selection criteria are listed in Table II. The calculated signal and SM background cross sections after all these selections, are presented in Table III. In Selection1, one isolated lepton ($e^+$ or $\mu^+$ ) is identified and used as trigger. In addition, $W^+$ boson is reconstructed by taking the missing transverse momentum as the neutrino momentum, and is fixed the longitudinal component by $M_W$. The most of the contribution to background cross section comes from jets of light quarks and gluon. In Selection2, I require $\left|M_{j\nu \ell}-M_{u_4}\right|\:<\:2\Gamma_{Tot}$, where $\Gamma_{Tot}$ is given in Fig. 2a. With this cut, the background is decreased by more than a factor of $100$. In Selection3, $b$-quark jet is tagged with efficiency of $56\%$, and $1\%$ of light jets (for $u$, $d$, $s$, $\bar{u}$, $\bar{d}$, $\bar{s}$ and $g$) and $10\%$ of $c$-quark jets are mistagged as b-jet in addition to Selection2. These cuts give further decrease on SM background cross section by almost a factor of $100$. In total, cut selections decrease background by a factor of $10^5$ with respect to no-cut case. The calculated signal and background cross sections, are plotted in Fig. 5 as a function of the reconstructed $b \nu \ell$ invariant mass. It is drawn for some $\left|V_{{u_4}b}\right|$ and mass values of the fourth generation quark for illustration.

\begin{figure}
\includegraphics{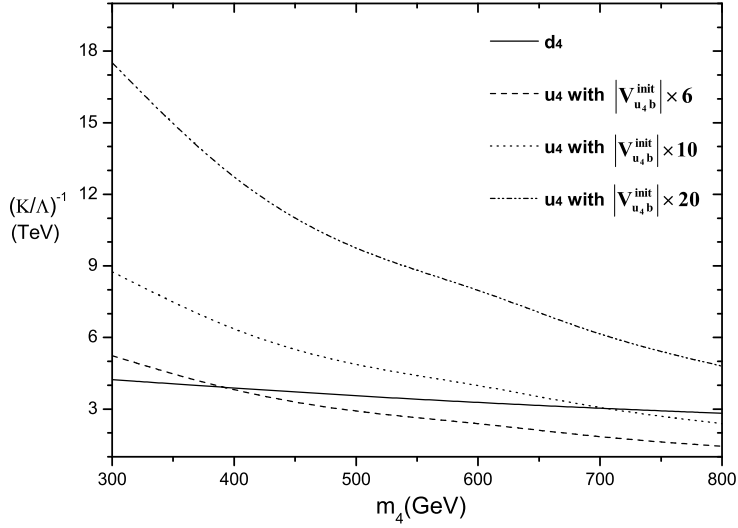}
\caption{Observation reach at $3\sigma$ for anomalous coupling strength as a function of the fourth generation quark mass for $pp\rightarrow u_4X\rightarrow b\nu \ell^{+} X$ and $p p \rightarrow d_4 X\rightarrow b\gamma X$ processes at the LHC.\protect\label{fig6}}
\end{figure}

\section{Conclusion}

The statistical significance (SS) values, evaluated from $SS=(\sigma_{S}/\sqrt{\sigma_B})\sqrt{L_{int}}$, where $L_{int}$ is the integrated luminosity of the collider, for both signal processes at the LHC with integrated luminosity of 100 fb$^{-1}$ are presented in Table IV. Achievable values of anomalous coupling strength as a function of the fourth generation quark mass for processes under consideration are shown in Fig. 6. For both processes $SS\geq3$ is taken as an observation criterion. One can see that values as low as 0.24 (0.35) TeV$^{-1}$ are reachable for $(\kappa/\Lambda)$ at $m_4=300\: (800)$ GeV for down type quark "anomalous production-anomalous decay" group process (as mentioned above). While anomalous interaction of the fourth generation up type quark is not observable with "anomalous production-SM decay" type process at $\left| V^{init}_{u_4 b}\right|$, it becomes better below 400 GeV mass of $u_4$ for $\left| V^{init}_{u_4 b}\right|\times 6$. $pp\rightarrow u_4X\rightarrow b\nu \ell^{+} X$ process for $\left| V^{init}_{u_4 b}\right|\times 10$ gets more observable below 700 GeV mass of fourth generation quark compare to $pp\rightarrow d_4X\rightarrow b\gamma X$ process. For $\left| V^{init}_{u_4 b}\right|\times 20$, reachable values for $(\kappa/\Lambda)$ are 0.057 (0.21) TeV$^{-1}$ at $m_4=300\: (800)$ GeV for up type quark "anomalous production-SM decay" group process.

As a result of this study, possible anomalous interactions of the fourth generation quarks at the given range of the quark mass, $\left| V_{u_4 b}\right|$ and $(\kappa/\Lambda)$ will be observed or excluded at the LHC in a few years.

\begin{acknowledgments}

I would like to thank A. K. \c{C}ift\c{c}i, S. Sultansoy and G. Unel for many helpful conversations and discussions, and acknowledge for support from the Turkish Scientific and Technical Research Council (TUBITAK) BIDEB-2218 grant. This work was also supported in part by the Turkish Atomic Energy Authority (TAEA) via "Turkish-ATLAS project".
 
\end{acknowledgments}


\begin{thebibliography}{0}

\bibitem{Yao06} W.-M. Yao {\it et al.}, {\it Journal of Physics G} {\bf 33}, 1 (2006).

\bibitem{He01} H.-J. He, N. Polonsky, S. Su, {\it Phys. Rev. D} {\bf 64}, 053004 (2001).

\bibitem{Novikov02} V. A. Novikov {\it et al.}, {\it Phys. Lett. B} {\bf 529}, 111 (2002).

\bibitem{Kribs07} G. D. Kribs, T. Plehn, M. Spannowsky, T. M. P. Tait, {\it Phys. Rev. D} {\bf 76}, 075016 (2007).

\bibitem{Sultansoy07} S. Sultansoy, AIP Conf. Proc. {\bf 899}, 49 (2007). 

\bibitem{Fritzsch87} H. Fritzsch, {\it Phys. Lett. B} {\bf 184}, 391 (1987).

\bibitem{Datta93} A. Datta, {\it Pramana} {\bf 40}, L503 (1993). 

\bibitem{Celikel95} A. \c{C}elikel, A. K. \c{C}ift\c{c}i, S. Sultansoy, {\it Phys. Lett. B} {\bf 342}, 257 (1995).

\bibitem{CDF07} CDF Collaboration, CDF Conf. Note, 8495 (2007).

\bibitem{Datta94} A. Datta, S. Raychaudhuri, {\it Phys. Rev. D} {\bf 49}, 4762 (1994).

\bibitem{Atag96} S. Atag, A. \c{C}elikel, A. K. \c{C}ift\c{c}i, S. Sultansoy, U. O. Yilmaz, {\it Phys. Rev. D} {\bf 54}, 5745 (1996).

\bibitem{Ciftci05} A. K. \c{C}ift\c{c}i, R. \c{C}ift\c{c}i , S. Sultansoy,  {\it Phys. Rev. D} {\bf 72}, 053006 (2005).

\bibitem{Chanowitz79} M. S. Chanowitz, M. A. Furman, and I. Hinchliffe, {\it Nucl. Phys. B}{\bf 153}, 402 (1979).

\bibitem{Sultansoy98} S. Sultansoy, {\it Turk. J. Phys.} {\bf 22}, 575 (1998).

\bibitem{ATLAS99} ATLAS: Detector and physics performance technical design report. Vol.~2, p.~519 CERN-LHCC-99-15, ATLAS-TDR-15, May 1999. 

\bibitem{Arik99} E. Ar{\i}k {\it et al.}, {\it Phys. Rev. D} {\bf 58}, 117701 (1998).

\bibitem{Holdom06} B. Holdom, {\it JHEP} {\bf 08}, 076 (2006).

\bibitem{Holdom07a} B. Holdom, {\it JHEP} {\bf 03}, 063 (2007).

\bibitem{Holdom07b} B. Holdom, {\it JHEP} {\bf 08}, 069 (2007).

\bibitem{Fritzsch99} H. Fritzsch, D. Holtmannspotter, {\it Phys. Lett. B} {\bf 457}, 186 (1999).

\bibitem{Arik03a} E. Ar{\i}k, O. Cak{\i}r, S. Sultansoy, {\it Europhys. Lett.} {\bf 62}, 332 (2003). 

\bibitem{Arik03b} E. Ar{\i}k, O. Cak{\i}r, S. Sultansoy, {\it Phys. Rev. D} {\bf 67}, 035002 (2003).

\bibitem{Ciftci07a} A. K. \c{C}ift\c{c}i {\it et al.}, AIP Conf. Proc. 899, 227 (2007).

\bibitem{Ciftci07b} A. K. \c{C}ift\c{c}i {\it et al.}, AIP Conf. Proc. 899, 191 (2007).

\bibitem{Ciftci08a} A. K. \c{C}ift\c{c}i, R. \c{C}ift\c{c}i, S. Sultansoy, {\it Phys. Lett. B} {\bf 660}, 534 (2008). 

\bibitem{Ciftci08b} A. K. \c{C}ift\c{c}i {\it et al.}, {\it Mod. Phys. Lett. A} {\bf 23}, 1047 (2008). 

\bibitem{Rizzo97} T. G. Rizzo, {\it Phys. Rev. D} {\bf 56}, 3074 (1997).

\bibitem{Boos04} E. Boos {\it et al.} (CompHEP Collaboration), Nucl. Instrum. Meth. A {\bf 534}, 250 (2004). 

\bibitem{Pumplin02} J. Pumplin {\it et al.}, JHEP {\bf 0207}, 012 (2002) [hep-ph/0201195].

\end{thebibliography}
\end{document}